# Do optical quarks exist in free space: scalar treatment?

Alexander V. Volyar

*General Physics Department, Taurida National University, Simferopol, Ukraine*
volyar@crimea.edu

We have considered new type of singular beams with fractional topological charges that were called the optical quarks possessed rather unique properties, their topological charges being half of the integer order. There are four types of optical quarks: even and odd with opposite signs of the topological charges. The sums or difference of the even and odd quarks form the standard vortex (or non-vortex) beams with the integer order topological charges. All quarks in the same beam annihilate, the beam vanishes. The analysis of the quark angular spectra showed that the optical quarks are structurally unstable field forms under the free-space propagation. We have analyzed their propagation properties for different types of the beam envelope including the symmetric beam array with discrete optical quarks. We have discussed the properties of the possible structurally stable quark forms and medium capable to maintain optical quarks.

## I. Introduction

As far back as in the beginning of the 90$^{th}$ M. Soskin at al [1,2] have wondered at a "strange" behavior of the simplest singular beams with fractional topological charge. It turns out that the inherent property of such beams was that the initial field distribution did not recovered when propagation at any beam length while the optical vortex with a fractional topological charge did not nucleated at any beam cross-section. Indeed, the broken axial symmetry of the beam does not permit to reconstruct the propagation field. The immediate inference was that the vortex beams with a fractional topological charge cannot exist in principle. Although such a simple statement does not need at all a strong confirmation the next paper [3] by Berry has stimulated the heated discussion. M. Berry considered the diffraction process of a Gaussian beam by a spiral phase plate with the fractional phase step. The evolution of the diffracted beam manifests itself in the form of the beam fracture with chains of singly charged optical vortices. However, the major point was that the beam could carry over a fractional orbital angular momentum (OAM). The avalanche of the subsequent papers surpassed all imaginations (see e.g. [4-10] and references therein). The detailed analysis showed that the fractional optical vortex splits into an infinite series of integer-order vortices while OAM of the beam is defined by contribution of the integer-order optical vortices. Although at the first sight it seems that the fractionalizing of the beam OAM contradicts the foundations of quantum mechanics the authors of the paper [11] showed the mixed stated of photons to be able to carry over the fractional OAM. At the same time, in according to the paper [3] the fractional-vortex beam must be inevitably destroyed when propagating because of different phase velocities of the partial elementary beams in its content. Nevertheless, the recent papers [12, 13] have demonstrated the space invariant beams with the fractional OAM and, in particular, with the fractional optical vortices [12] – the so-called erf-G beams. It is important to stress that with the pattern of the fractional edge dislocations we meet also in the Ahoronov –Bohm effect and in the surface waves in the swimming pool above the sink [14]
Such unusual properties of the fractional-vortex beams compel to peer more attentively into the structure of the space-invariant fractional vortex beams.

The aim of the paper is to analyze the structural features of the elementary fractional-vortex beams in free space.

## II. The erf-G beams as anticipation of the optical quarks

### *II.1 Fundamental quark properties*

In the paper [12] we have shown that the error function-Gaussian beams (erf-G beams) bearing the optical vortices with $l = \pm 1/2$ topological charge are of the strong solutions to the vector paraxial wave equation and refer to the set of the so-called standard paraxial beams such as Hermite-Gaussian (HG), Laguerre-Gaussian (LG), Bessel-Gaussian (BG) etc beams with a complex argument. In contrast to the usual standard beams (HG,L,G, BG etc) the erf-G beams have the non-factorizing form that is the azimuthal $\varphi$ and radial $r$ variables in them are not separated. The scalar erf-G beam can be written in the form

$$\Psi_s = -\frac{2i\sqrt{\pi}\, e^{is\varphi/2}}{\Re} N G \left\{ e^{-\Re^2/2} \mathrm{erf}\left(i\Re \sin\frac{\varphi}{2}\right) + s\, e^{\Re^2/2} \mathrm{erf}\left(\Re \cos\frac{\varphi}{2}\right) \right\}, \qquad (1)$$

where $\mathrm{erf}(x)$ stands for the error function, $\Re = \sqrt{2\dfrac{K r}{\sigma}}$, $G = \exp\left(-\dfrac{r^2}{w_0^2 \sigma}\right)/\sigma$, $N = \exp\left(-\dfrac{K^2 w_0^2}{4\sigma}\right)$, $\sigma = 1 - i z / z_0$, $z_0 = k w_0^2 / 2$, $w_0$ is the radius of the beam waist at $z=0$, $k$ is the wavenumber, $s = \pm 1$, the free parameter $K$ can be arbitrary value including the complex one. The field distribution (1) depends on the free parameter $K$. Some intensity and phase distributions in Fig.1 illustrate variations of the beam structure for the real and imaginary $K$-parameter at the initial plane z=0. For the case of the pure imaginary $K$ parameter the field represents the complex set of singly charged optical vortices evolving the complex way when propagating. When the $K$ parameter is a pure real value, the intensity distribution has the C-like shape. The ray along $\varphi = 0$ is something like a broken edge dislocation that matches two edges of the wave function (1) with different phases in the one-half optical vortex. Even very small shift from the initial plane breaks the spiral-like phase structure in Fig.1d resulting in the phase smoothing [12].

Near the beam axis where $K r$ is a very small $K r \ll 1$ the wave function (1) of the erf-G beam can be presented as

$$\Psi_s \approx 2\sqrt{\pi}\, e^{is\varphi/2} N G \left\{ e^{-\Re^2/2} \sin\frac{\varphi}{2} - i s\, e^{\Re^2/2} \cos\frac{\varphi}{2} \right\}. \qquad (2)$$

But two terms in eq.(2) have much to do with the forms of the fractional-vortex beams proposed by Soskin et al in the paper [1]. Let us write the generalized form of such wave constructions at the initial plane z=0 and outline their basic properties.

$$Q_{ev}^{+,m} = \cos\left(m\frac{\varphi}{2}\right) e^{im\frac{\varphi}{2}} F(r), \qquad (3)$$

$$Q_{od}^{+,m} = -i \sin\left(m\frac{\varphi}{2}\right) e^{im\frac{\varphi}{2}} F(r), \qquad (4)$$

$$Q_{ev}^{-,m} = -\cos\left(m\frac{\varphi}{2}\right) e^{-im\frac{\varphi}{2}} F(r), \qquad (5)$$

$$Q_{od}^{-,m} = -i \sin\left(m\frac{\varphi}{2}\right) e^{-im\frac{\varphi}{2}} F(r), \qquad (6)$$

where $F(r)$ is the radial envelope of the standard paraxial beam, $m = 2m' + 1$ is of the odd numbers.

Arbitrary non-vortex beam can be presented as the superposition of the wave elements (3)-(6):

$$F(r) = Q_{ev}^{+,m} + Q_{od}^{+,m} \quad \text{or} \quad F(r) = -\left(Q_{ev}^{-,m} + Q_{od}^{-,m}\right). \qquad (7)$$

Similar to that we can define arbitrary paraxial vortex beam with the odd topological charge
$$F(r)e^{im\varphi} = Q_{ev}^{+,m} - Q_{od}^{+,m} \text{ and } F(r)e^{-im\varphi} = Q_{od}^{-,m} - Q_{ev}^{-,m}. \tag{8}$$

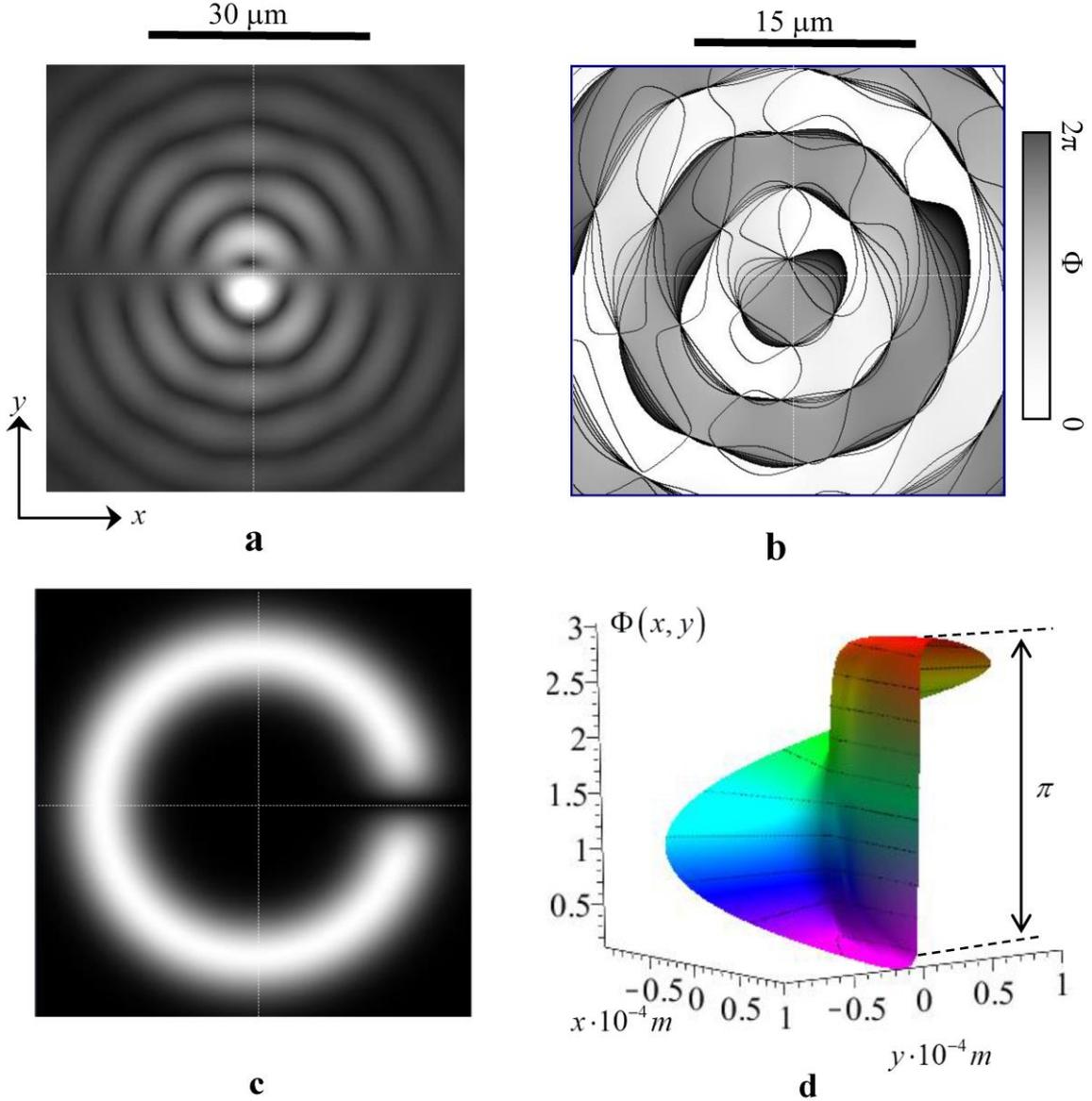

Fig.1 Intensity (a,c) and phase (b,d) distributions in erf-G beam with $w_0 = 35\,\mu m$, for (a,b) , $K_\perp = i\,6\cdot 10^5 m^{-1}$ at the initial plane $z = 0$ and for (c,d) $K_\perp = 6\cdot 10^5 m^{-1}$ at the initial plane $z = 0$

Correspondingly, beams with the edge dislocations are written as
$$F(r)\cos(m\varphi) = \{Q_{ev}^{+,m} - Q_{od}^{+,m} + Q_{od}^{-,m} - Q_{ev}^{-,m}\}/2, \tag{9}$$
$$F(r)\sin(m\varphi) = \{Q_{ev}^{+,m} - Q_{od}^{+,m} - Q_{od}^{-,m} + Q_{ev}^{-,m}\}/2i. \tag{10}$$

However, the vortex beam of the high orders is the unstable ones against slight perturbations we will focus our attention later on only on the simplest vortex beams with $m = \pm 1$.

One more basic property is that the sum all wave constructions (3) –(6) vanishes:
$$Q_{ev}^{+,m} + Q_{od}^{+,m} + Q_{ev}^{-,m} + Q_{od}^{-,m} = 0. \tag{11}$$

In analogy with the Gell-Mann quark model of the hadrons [15] we have called such wave constructions the *optical quarks*. In this way *the wave constructions $Q_{od}^{-,m}, Q_{ev}^{-,m}$ can be treated as anti-quarks. The equations (7) and (8) can be read as the superposition of two even and odd quarks or anti-quarks forms the non-vortex beam while their difference is the vortex-beam. At*

*the same time, the occurrence of all quarks and anti-quarks results in their total annihilation* (see eq.(11)).

Typical intensity and phase distributions in the optical quark are shown in Fig.2. It draws attention the distinct phase step $\pi$ along the corresponding rays including the axis $r=0$,

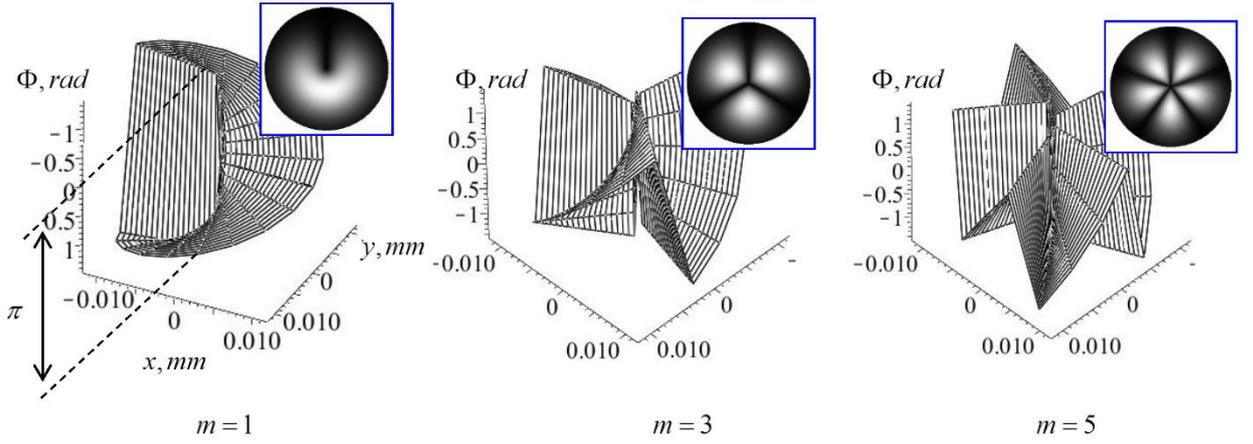

Fig.2 The intensity $I$ and phase $\Phi$ distribution of the optical quark $Q_{ev}^{+,1}$ with the standard envelope $F(r) = (r/w_0)^m G$

### II.2 Angular spectrum of the simplest optical quarks

The propagation properties of the optical quarks are defined by their angular spectrum functions

$$\Psi(k_p,\phi) = \frac{k}{2\pi}\int_0^\infty r\,dr \int_0^{2\pi} d\varphi\, \Psi(r,\varphi) e^{-ik_p r\cos(\varphi-\phi)}, \quad (12)$$

where $k_p$ and $\phi$ are the radial and azimuthal coordinates in the **k**-space, $r$ and $\varphi$ are the polar coordinates.

At first we will analyze the angular spectrum of the optical quark with BG envelope at the initial plane z=0 in the form

$$Q_{ev}^{+,1} = \cos\frac{\varphi}{2} e^{i\frac{\varphi}{2}} J_1(Kr) e^{-r^2/w_0^2}. \quad (13)$$

Since $\cos\frac{\varphi}{2} e^{i\frac{\varphi}{2}} = \frac{e^{i\varphi}+1}{2}$, the spectral integral (12) can be presented as the sum of two functions

$$\Psi(k_p,\phi) = \Psi_1(k_p,\phi) + \Psi_2(k_p,\phi), \quad (14)$$

The form of the $\Psi_{1,2}$ functions is defined by the $K$-parameter. If the $K$ is a real value we obtain

$$\Psi_1(k_p,\phi) = -iz_0 e^{i\phi} \exp\left\{-\frac{(K^2+k_p^2)w_0^2}{4}\right\} I_1\left(\frac{k_p K w_0^2}{2}\right), \quad (15)$$

where $I_m(x)$ is the modified Bessel function of the first kind and the *m*-th order.

For the imaginary value of the $K$-parameter we find

$$\Psi_1(k_p,\phi) = z_0 e^{i\phi} \exp\left\{\frac{(K^2-k_p^2)w_0^2}{4}\right\} J_1\left(\frac{k_p K w_0^2}{2}\right) \quad (16)$$

because of $I_n(ix) = i^n J_n(x)$, $J_m(x)$ is the Bessel function of the first kind and the *m*-th order.

The second integral is

$$\Psi_2(k_p) = z_0 \frac{k_p w_0}{2} \sum_{m=0}^{\infty} \frac{\Gamma(m+3/2)}{(m!)^2} \left(-\frac{k_{p-}^2 w_0^2}{4}\right)^m {}_2F_1\left(-m, -(m+1); 1; \pm\frac{K^2}{k_p^2}\right), \quad (17)$$

where the sign $(+)$ refers to the real K, while the sign $(-)$ is associated with the imaginary $K$-parameter. ${}_2F_1$ stands for the confluent hypergeometric function and $\Gamma(x)$ is the Gamma function.

Typical amplitude and phase distributions in the optical quark with the Bessel-Gaussian envelope both for the real and the imaginary $K$-parameters are shown in Fig.3. First of all. these are the complex angular spectra with a series of main directions of the wave propagation that are asymmetrically positioned. The second, there are topological dipoles (two opposite charged phase screw singularities) in the **k**-space phase distributions.

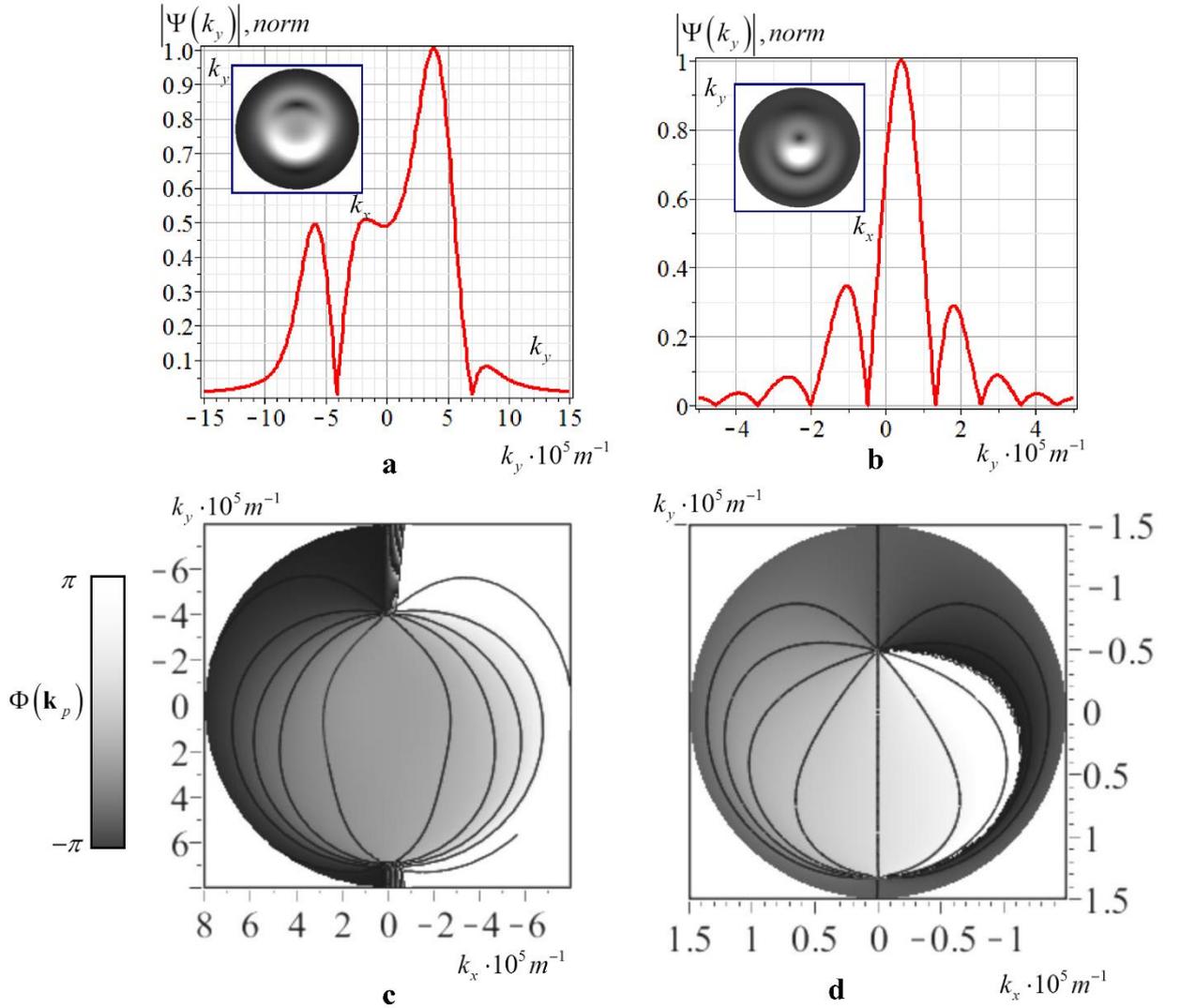

Fig.3 Amplitude $|\Psi(k_p, \phi)|$ (a,b) and phase $\Phi(\mathbf{k}_p) = \arg(\Psi(k_p, \phi))$ (c,d) for the optical quark $Q_{ev}^{+,1}$ with the Bessel-Gaussian envelope; (a,c) $K = i5 \cdot 10^5 m^{-1}$, $w_0 = 10^{-5} m$, (b,d)

Similar to that we can construct the angular spectrum for the optical quarks with the simplest LG envelope at the $z=0$ plane:

$$\Psi(r, \varphi) = \cos\frac{\varphi}{2} e^{i\frac{\varphi}{2}} \left(\frac{r}{w_0}\right) e^{-r^2/w_0^2}. \quad (18)$$

Now we obtain two spectrum functions in eq, (11) in the form

$$\Psi_1(k_p) = \frac{z_0\sqrt{\pi}}{2}\left[\left(1-\frac{k_p^2 w_0^2}{4}\right)I_0\left(\frac{k_p^2 w_0^2}{8}\right)+\frac{k_p^2 w_0^2}{4}I_1\left(\frac{k_p^2 w_0^2}{8}\right)\right]e^{-\frac{k_p^2 w_0^2}{8}}, \quad (19)$$

$$\Psi_2(k_p,\phi) = \frac{k_p w_0 z_0 e^{i\phi}}{2} e^{-\frac{k_p^2 w_0^2}{4}}. \quad (20)$$

Typical features of the angular spectrum of the optical quark with the LG envelope are shown in Fig.4. As same as in the case of the quark with the BG envelope we observe the complex angular spectrum with the topological dipole. The presence of the energy flux along the axis $k_p=0$ points out the transformation of the quark structure along the beam axis. Besides, the topological dipole in the spectral function presupposes dislocation reactions in the wave structure when propagating the beam, the above features being, by all appearances, inherent in either standard type of the quark envelope in free space.

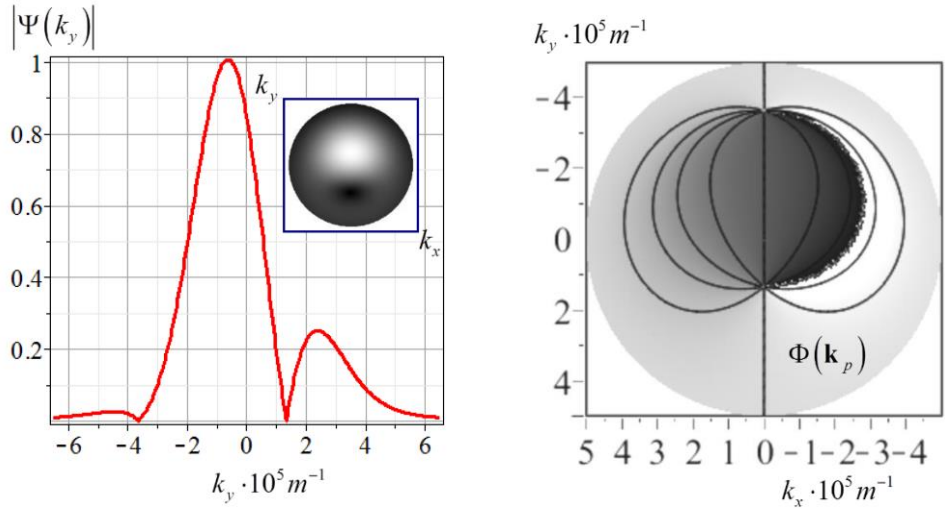

Fig.4 Amplitude $|\Psi(\mathbf{k}_p)|$ and phase $\Phi(\mathbf{k}_p)$ distributions in the angular spectrum of the optical quark $Q_{ev}^{+,1}$ with the LG envelope

### II.3 The free space propagation of the optical quarks

The detailed propagation process of the paraxial beams can be described in the frameworks of the diffraction integral

$$\Psi(r,\varphi,z) = \frac{C}{z}\int_0^\infty dr' \int_0^{2\pi} e^{-ik\sqrt{(x-x')^2+(y-y')^2+z^2}} \Psi(r,\varphi')d\varphi', \quad (21)$$

where $C$ is constant. For the optical quark $Q_{ev}^{+,m}$ we have

$$\Psi(r,\varphi,z) = \frac{C}{z}\int_0^\infty dr' \int_0^{2\pi} e^{-ik\sqrt{(x-x')^2+(y-y')^2+z^2}} \left\{\cos\left(m+\frac{1}{2}\right)\varphi' e^{i\left(m+\frac{1}{2}\right)\varphi'} r'^m e^{-\frac{r'^2}{w_0^2}}\right\}d\varphi' \quad (22)$$

So that in the paraxial approximation for the wave zone we obtain

$$Q_{ev}^{+,m}(r,\varphi,z) = \frac{Ce^{-ikz-\frac{ikr^2}{2z}}}{2z}\int_0^\infty r'^2 dr' e^{-\frac{ikr'^2}{2z}} e^{-\frac{r'^2}{w_0^2}}\int_0^{2\pi}(1+e^{im\varphi'})e^{i\frac{krr'}{z}\cos(\varphi-\varphi')}d\varphi' = \frac{Ce^{-ikz}}{2}(\Psi_{m,1}+\Psi_{m,2}), \quad (23)$$

with

$$\Psi_{1,m} = \frac{\Gamma\left(\frac{m+1}{2}\right) z^{\frac{m}{2}+1} e^{\frac{kz_0}{4z(z+iz_0)}r^2} e^{-ik\frac{r^2}{2(z+iz_0)}}}{k^2 r^2 (z+iz_0)^{\frac{m}{2}}} \left\{ mM_{\frac{m}{2},\frac{1}{2}}\left(\frac{kz_0}{2z(z+iz_0)}r^2\right) - \right.$$
$$\left. -(2+m) M_{\frac{2+m}{2}}\left(\frac{kz_0}{2z(z+iz_0)}r^2\right)\right\} \quad , \quad (24)$$

$$\Psi_{m,2} = \left(\frac{z_0}{z+iz_0}\right)^m \frac{w_0^2}{z+iz_0}\left(\frac{r}{w_0}\right)^m e^{im\varphi} e^{-ik\frac{r^2}{2(z+iz_0)}}, \quad (25)$$

$M_{\mu,\nu}(x)$ stands for the Whittaker function.

For the simplest case m=1 we find

$$\Psi = \frac{\sqrt{z\pi}\, w_0^2 e^{\frac{kz_0}{4z(z+iz_0)}r^2}}{4(z+iz_0)^{\frac{3}{2}}} \left\{ \left(1 - \frac{kz_0}{2z(z+iz_0)}r^2\right) I_0\left(\frac{kz_0}{4z(z+iz_0)}r^2\right) + \right.$$
$$\left. + \frac{kz_0}{2z(z+iz_0)} r^2 I_1\left(\frac{kz_0}{4z(z+iz_0)}r^2\right)\right\} e^{-ik\frac{r^2}{2(z+iz_0)}} \quad . \quad (26)$$

The wavefront transformations at the wave zone for the optical quarks with simples LG envelope are shown in Fig.5. A slight shift of the observation plane along the beam axis is accompanied by nucleating a chain of the singly charged optical vortices with opposite topological charges that involve when increasing the length. At far field the field structure is simplified (Fig. 6), the vortex chains are extruded at infinity the distance between them increases so that near the beam axis we observe only one singly charged optical vortex.

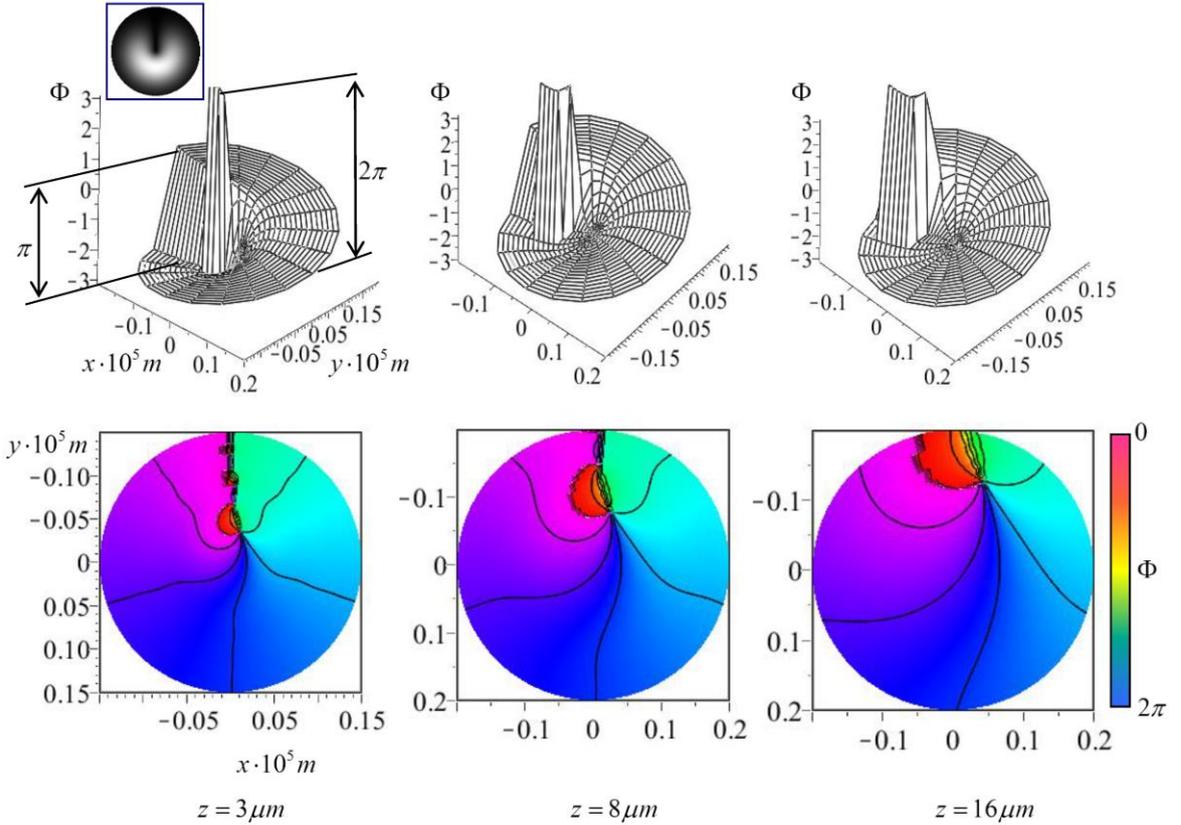

$z = 3\mu m \qquad z = 8\mu m \qquad z = 16\mu m$

Fig.5 Wavefront $\Phi(r,\varphi,z)$ evolution of the optical quark $Q_{ev}^{+,1}$ with $w_0 = 20\,\mu m$ along the z direction in the wave zone

Thus, the inner structure of the optical quark with a continuous phase distribution at the beam cross-section is such that either standard envelope of the wave construction is inevitably accompanied by the quark destruction. However, there is a lot of other non-standard forms of the beam envelopes for paraxial beams that might nevertheless provide a structural stability of the optical quarks. One of them is discrete optical vortices [16,17].

### III. Discrete optical quarks

A new type of screw wavefront singularities is the discrete optical vortices that are shaped in the symmetrical beam arrays [16-19]. The beam array can be constructed in such a way that the field amplitude near the axis vanishes that can essentially weaken the optical quark destruction. Besides, variation of the array parameters permits it to be transformed into the so-called spiral beam [17] – that is into the structurally stable wave construction whose shape is recovered up to the scale and rotation when propagating.

In our consideration we will hold later on the approach of the paper [19]. Let us consider the symmetric array of Gaussian beams shown in Fig. 7. Each Gaussian beam is located at the vertices of the regular polygon ($x'_n, y'_n, z$ - local coordinates) at the distance $r_0$ from the origin of the global reference frame. The $z$-axis of the each local beam is tilted at the angle $\alpha$ at the $(y'_n, z)$ plane - ($x_n = x'_n, y_n, z_n$ - coordinates).

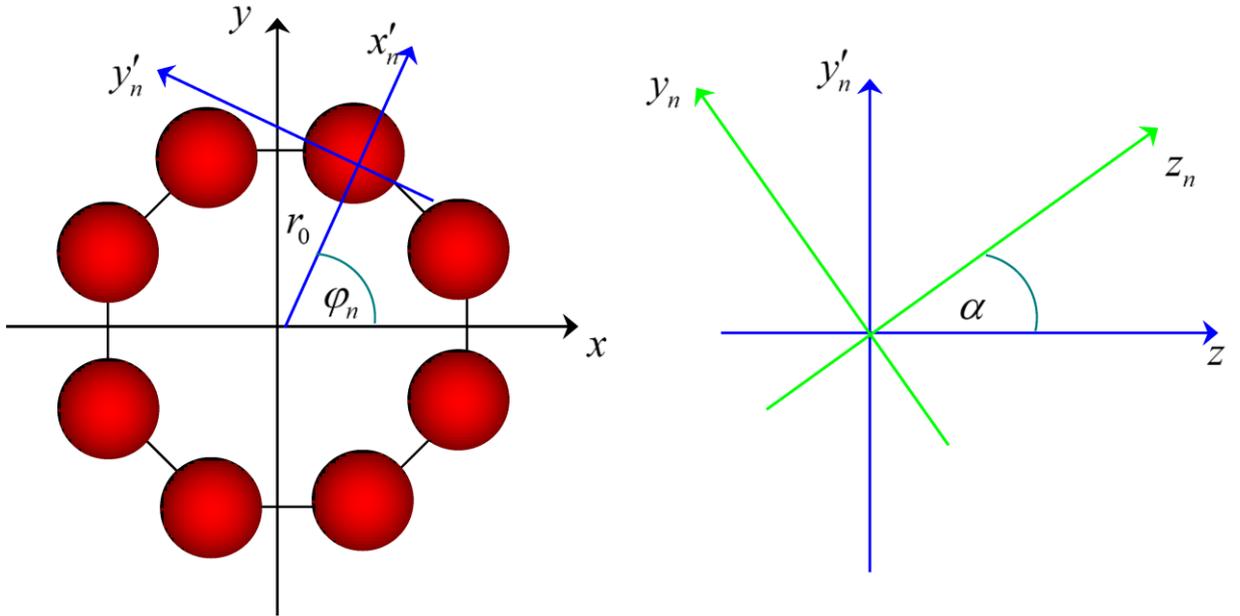

Fig.7 Geometry of the beam array

The beam coordinates are connected with the coordinates in the global referent frame $(x, y, z)$ as
$$x_n = r\cos(\varphi+\varphi_n)+r_0, \quad y_n = r\sin(\varphi+\varphi_n)-\alpha z, \quad z_n = \alpha r\sin(\varphi+\varphi_n)+z, \quad (27)$$
where we used the paraxial approximation $\sin\alpha \approx \alpha$, $r$ and $\varphi$ are the polar coordinates of the global referent frame, $\varphi_n = n\frac{2\pi}{N}$, $N$ is a number of local beams in the array. Then the local phase $\Delta_{n,q} = \varphi_n q$ and amplitude $f_{n,q}$ with $q$ is arbitrary real number are given to each beam. Thus the scalar field of the beam array can be obtained in the form
$$\Psi_q = \frac{1}{N}\sum_{n=0}^{N-1} f_{n,q} e^{i\Delta_{n,q}} \frac{\exp\left\{-ik\frac{x_n^2+y_n^2}{2Z_n}\right\}}{Z_n/(iz_0)} e^{-ikz_n}, \qquad (27)$$

where $Z_n = z_n + i z_0$.

When the number $q$ is integer one and $f_{n,q} = 1$ the total pass-by of the beam axis changes beam phase by the value $2\pi q$. The beam gets the optical vortex with the topological charge equal to $q$. But the phase at the local beams changes step by step rather than continuously. Such a strange vortex beam came to be called the discrete optical vortex [16]. The same situation occurs in the case when $q = m/2$ where $m$ is an odd number but the phase changes step by step at the total value $\pi m n / N$ whereas the amplitude is $\cos\left(\dfrac{\pi m}{N} n\right)$ or $\sin\left(\dfrac{\pi m}{N} n\right)$. We called them the *discrete optical quarks*.

The simples case of the discrete optical quark formed by the beam array when all axes of the local beams parallel $(\alpha = 0)$ to the global axis $z$ and $f_{n,m} = \cos\left(\dfrac{\pi m}{N} n\right)$, $\Delta_{n,m} = \dfrac{\pi m}{N} n$ is shown in Fig,8 at the initial plane $z = 0$. In contrast to the ideal phase patterns of the quark in Fig.2 the phase step in this case has the non-uniform shape. Although far from the axis the phase step is $\pi$ the phase reaches gradually the value $\pi$ near the array axis. Besides we do not deal hear with the spiral beam and its structure therefore must change as the beam propagates. Indeed, Fig.9 shows that intensity distribution of the optical quark with $q = 5/2$ evolves very quickly along the array axis so that the beam gets a rather regular tracery of the singly charged optical vortices at far field.

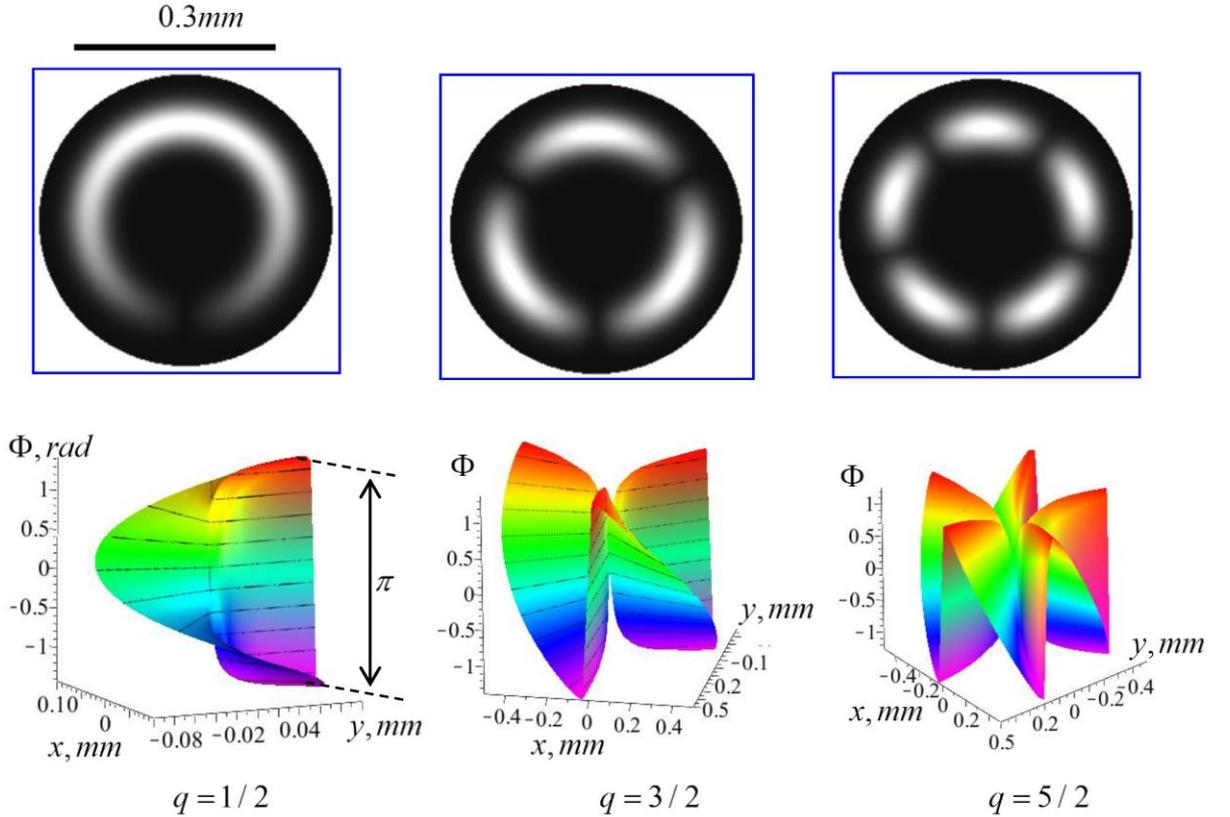

Fig.8 Intensity $I$ and phase $\Phi$ distributions in the discrete optical quarks with $N = 100$, $w_0 = 20\mu m$, $\alpha = 0$, $r_0 = 0.1 mm$ at $z = 0$

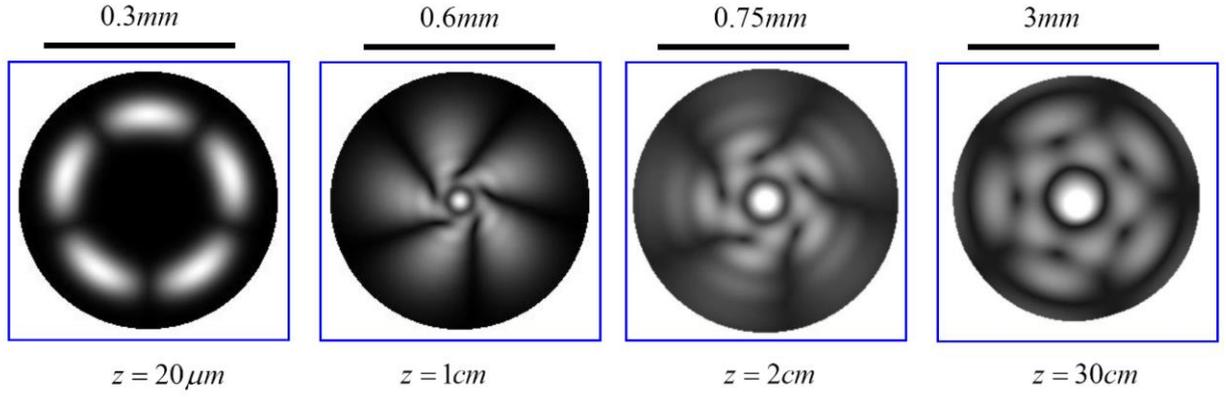

Fig.9 Space evolution of the intensity distribution in the discrete optical quark with $q = 5/2$, $N = 100$, $w_0 = 20 \mu m$, $\alpha = 0$, $r_0 = 0.1 mm$

On the other hand, under the condition $r_0 = \alpha z_0$ the beam array turns into the spiral beam at least for the integer $q$ [17]. It means that the tilt of the local beams is compensated by the diffraction process so that the field is recovered up to the scale and rotation. It is the case that is shown in Fig.10. We observe here the cylindrical phase plate with $\Phi = 0$ in vicinity of the array axis. The corresponding $\pi$-steps are shifted but have the regular form of the step (in contrast to that in Fig.8. The space evolution of the optical quark shown in Fig.11 is of the destruction process. The shape of the wavefront is gradually deformed, the spiral beam loses it's the self-recovered properties for the fractional q-index, the discrete optical quark is broken down. At the same time, the sum of the even and odd quarks remains invariant structure.

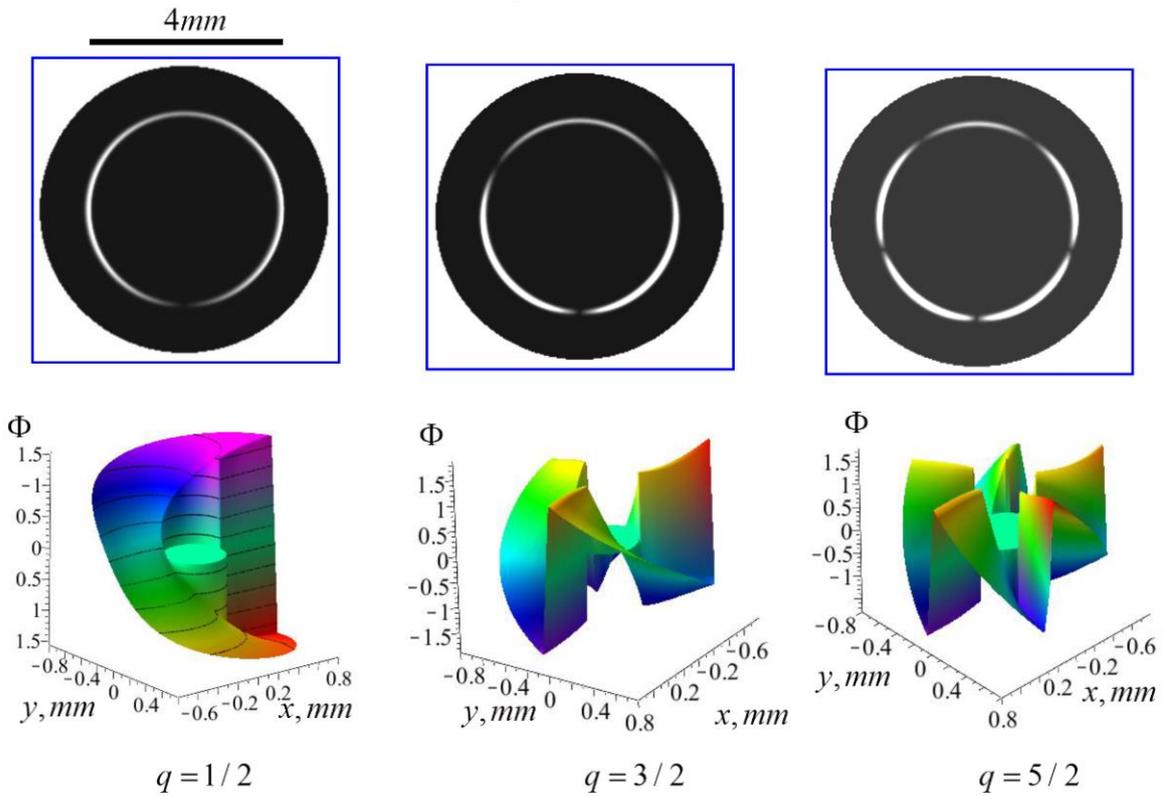

Fig.10 Discrete quarks in the spiral beam with $N = 100$, $r_0 = 4mm$, $\alpha = 0.0008$, $w_0 = 20\mu m$ at $z = 0$

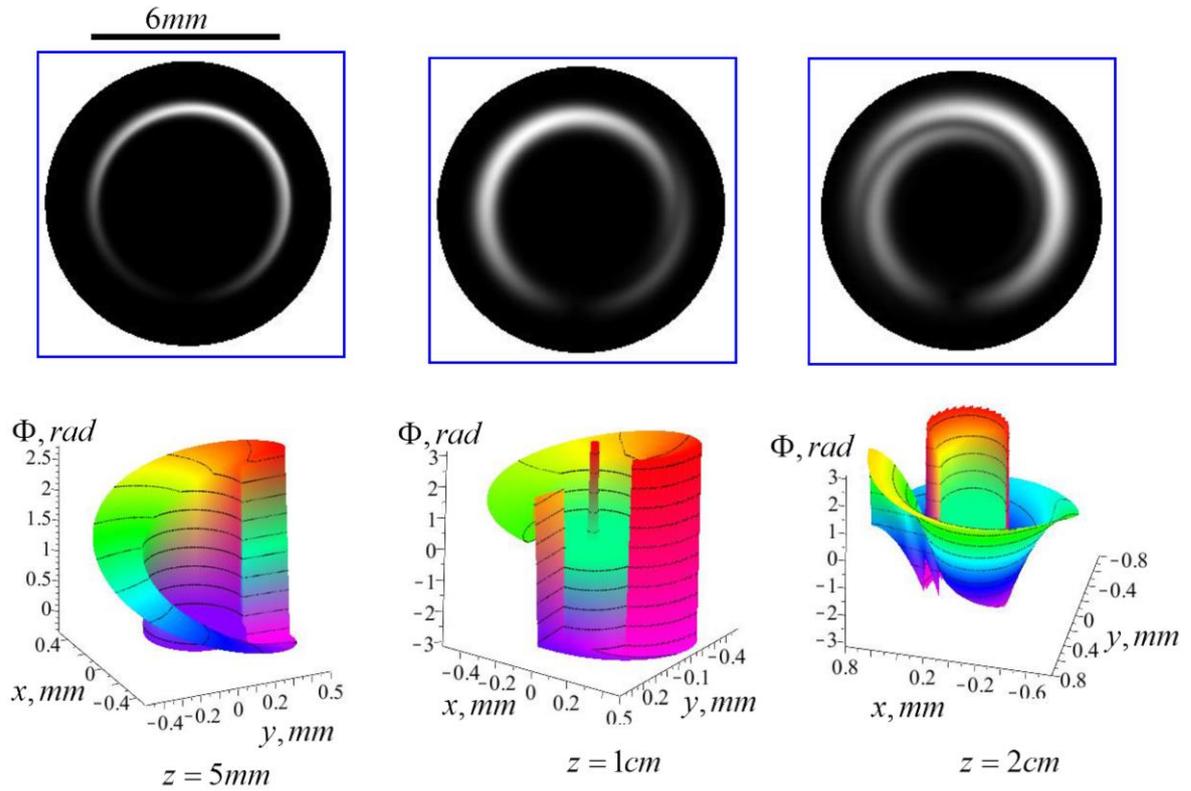

Fig.11 Space evolution of the optical quark in the spiral beam, $q=1/2$

## IV. Conclusions

We have considered new type of singular beams with fractional topological charges that were called the optical quarks possessed rather unique properties, their topological charges being half of the integer order. There are four types of optical quarks: even and odd with opposite signs of the topological charges. The sums or difference of the even and odd quarks form the standard vortex (or non-vortex) beams with the integer order topological charges. All quarks in the same beam annihilate. The analysis of the quark angular spectra showed the presence of the spectral components directed along the beam axis. It means that the quark wave state is structurally unstable one under the beam propagation. We have analyzed the propagation properties of the optical beams with different types of standard envelope including the symmetric beam array both in arbitrary and the self-recovered (spiral beams) states. We revealed that the spiral beams with the half-order indices lose their self-recovered properties that cause the quark breaking down.

The destruction of the optical quarks in free space put definite restrictions onto the beam structure and the properties of the medium capable to preserve the optical quarks. From our point of view the structurally stable optical quark must be, first of al, the vector field whose circular polarized components have the half integer order topological charges differ by one unite. Besides the medium has to be birefringent properties where the tensor principle directions form the field with fractional topological index.